\documentclass{amsart}
\usepackage{graphicx}
\newtheorem{thm}{Theorem}[section]
\newtheorem{cor}[thm]{Corollary}

\newtheorem{prop}[thm]{Proposition}
\theoremstyle{definition}
\newtheorem{definition}[thm]{Definition}
\theoremstyle{remark}

\numberwithin{equation}{section}
\def\ds{\displaystyle}
\newcommand{\R}{\mathbb{R}}
\newcommand{\N}{\mathbb{N}}
\newcommand{\Z}{\mathbb{Z}}
\newcommand{\C}{\mathbb{C}}

\newcommand{\bea}{\begin{eqnarray*}}
\newcommand{\eea}{\end{eqnarray*}}
\newcommand{\bean}{\begin{eqnarray}}
\newcommand{\eean}{\end{eqnarray}}

\begin{document}

\title[Sturm-Liouville problem in Quantum calculus]{Sturm-Liouville problem in Quantum calculus}
\author{Ahmed Fitouhi}
\address{D\'epartement de Mathematiques\\ Facult\'e des Sciences de Tunis,
 1060 Tunis, Tunisia.}
 \email{Ahmed.Fitouhi@fst.rnu.tn}

\author{Akram NEMRI}
\address{D\'epartement de Mathematiques\\ Facult\'e des Sciences de Tunis \\ $1060$ Tunis, Tunisia }
\email{Akram.Nemri@fst.rnu.tn} 
\author{Meniar Haddad}
\address{ Facult\'e des Sciences de Tunis,
 1060 Tunis, Tunisia.}
\email{Meniar.Haddad@fst.rnu.tn}


\begin{abstract}
This paper aims to study the $q$-analogue of the Sturm Liouville
problem and to give an asymptotic behaviour at infinity for its
solution $\varphi$. Additionally, we establish an asymptotic
expansion of the $q$-Bessel function $j_{\alpha}$ for $\alpha
>-\frac{1}{2}$. We are not in situation to claim that our results
are new but they have the advantage to show that the method used by
Agranovich and Marchenko remain true.
\end{abstract}
\maketitle
\section{Introduction:}

In the classical spectral analysis (see \cite{am} , \cite{mar},...),
we denote by $L$ a linear differential operator of the second-order
given by on $[0,\infty[$ of the form \bean
Lu=\frac{d^2u}{dx^2}-p(x)u\eean where
$p(x)$ is a real function, continuous which is integrable on $[0,\infty[$.\\

 We note by $\varphi(x,\lambda^2)$ and $\theta(x,\lambda^2)$ the
solutions of \bean \label{ec}Lu=\lambda^2 u\eean with initial conditions\\

$\begin{array}{cc}
  \varphi(0,\lambda^2)=\sin\alpha,&\varphi'(0,\lambda^2)=-\cos\alpha;\\
  \theta(0,\lambda^2)=\cos\alpha, &\theta'(0,\lambda^2)=\sin\alpha, \\
\end{array}$\\

where $\lambda$ is an arbitrary positif real number and $\alpha$ is
an arbitrary real number.

It is known that solving (\ref{ec}) is equivalent to solving the
following volterra integral equation  \bean
u(x,\lambda^2)=\sin\alpha \cos(\lambda x)-\frac{\sin(\lambda
x)}{\lambda}\cos\alpha +\int_0^x\frac{\sin(\lambda(x-y))}{\lambda}
p(y) u(y,\lambda^2) dy \eean where $x\in[0,+\infty[$, $\lambda \in
\mathbb{R_+^*}$ and $p(x)$ is an continuous
integrable function on $[0,+\infty[$.\\
 Hence, for  all
$\lambda\geq\rho>0$, $\varphi(x,\lambda^2)$ is a bounded function
and have the asymptotic formulas \bean
\varphi(x,\lambda^2)=\mu(\lambda^2) \cos(\lambda
x)+\nu(\lambda^2)\sin(\lambda x)+O(1) \eean where \bean
\mu(\lambda^2)&=&\sin\alpha-\int_0^\infty\frac{\sin(\lambda
y)}{\lambda} p(y) \varphi(y,\lambda^2) dy,\\
\nu(\lambda^2)&=&-\frac{\cos\alpha}{\lambda}+\int_0^\infty\frac{\cos(\lambda
y)}{\lambda} p(y) \varphi(y,\lambda^2) dy. \eean  Similarly, we have
\bean \theta(x,\lambda^2)=\mu_1(\lambda^2) \cos(\lambda
x)+\nu_1(\lambda^2)\sin(\lambda x)+O(1) \eean where \bean
\mu_1(\lambda^2)&=&\cos\alpha-\int_0^\infty\frac{\sin(\lambda
y)}{\lambda} p(y) \theta(y,\lambda^2) dy,\\
\nu_1(\lambda^2)&=&\frac{\sin\alpha}{\lambda}+\int_0^\infty\frac{\cos(\lambda
y)}{\lambda} p(y)\theta(y,\lambda^2) dy. \eean In the present paper
we are concerned to give its $q$-analogue and study its asymptotic
behaviour at infinity.\\
This paper is organized as follows: in section 2, we present some
preliminaries results and notations that will be useful in the
sequel. Further it is natural to consider in section 3, the
asymptotic behaviour of $\varphi(x,\lambda^2;q^2)$ and
$\theta(x,\lambda^2)$ for  $\lambda \longrightarrow\infty$. the
fundamental result is given in the following theorem
\begin{thm}
For $\lambda$ in $\R_{q,+}$, we have:
\begin{equation}
\mu(\lambda^2 ; q^2)\nu_1(\lambda^2 ; q^2)- \nu(\lambda^2 ;
q^2)\mu_1(\lambda^2 ; q^2) = \frac{1}{q^{\frac{1}{2}} \lambda}.
\end{equation}
\end{thm}
Section 4, is devoted to finding precise asymptotic formulas of
$j_\alpha$: the $q$-Bessel function for large $\lambda$.
\section{Notations and preliminaries}
  We recall some usual notions and notations
used in the $q$-theory. Let $a$ and $q$ be real numbers such that $0
< q <1$. In all the sequel we suppose that and
$\ds\frac{\log(1-q)}{\log q} \in \Z$.\\ The $q$-shifted factorials
are defined by
\begin{eqnarray}
 (a ; q)_n &=& \prod^{n-1}_{k=0} (1 - aq^k)\quad ; n \in \N\backslash\{0\},\\
  (a; q)_0 &=&1,\\
 (a ; q)_\infty &=& \prod^{\infty}_{k = 0} (1 - aq^k)
\end{eqnarray}
and more generally:
\begin{eqnarray}
 (a_1,\cdots,a_r ; q)_n &=& \prod^r_{k=1} (a_k ; q)_n.
\end{eqnarray}
The basic hypergeometric series or $q$-hypergeometric series is
given for r , s integers by
$${}_r\phi_s(a_1 ,\cdots,a_r; b_1,\cdots,b_s ; q,x) =
\sum^\infty_{n=0} \frac{(a_1,\cdots,a_r;q)_n}{(b_1,\cdots,b_s
;q)_n(q,q)_n}[(-1)^n q^{\frac{n(n-1)}{2}}]^{1+s-r} x^n $$
 here $$r,s \in \N ; a_1,\cdots,a_r\in \C ; b_1,\cdots,b_s \in
\C \backslash \{q^{-k}, k \in \N\}$$ The $q$-derivative $D_{q,x} f$
of a function $f$ on an open interval is given by:
\begin{equation}\label{der}
 D_{q,x}f(x) = \frac{f(x)
- f(qx) }{(1-q)x},\quad x\neq 0
\end{equation}
and $(D_{q,x} f )(0)=f^{'}(0)$ provided $f^{'}(0)$ exist. The
$q$-shift operators are
\begin{eqnarray}
(\Lambda_{q,x} f)(x)&=& f(q x)\\
(\Lambda^{-1}_{q,x}f)(x)&=& f(q^{-1}x).
\end{eqnarray}
We consider the $q$-operator
\begin{equation}
\Delta_{q,x} =\Lambda^{-1}_{q,x}D^{2}_{q,x}.
\end{equation}
The $q$-Jackson integral from $0$ to a and from $0$ to $\infty$ are
respectively defined by
\begin{eqnarray}
\int_{0}^{a}f(x)d_q x &=&(1-q)a \sum_{n=0}^{\infty}f(aq^n )q^n \\
\int_{0}^{\infty}f(x)d_q x&=& (1-q)\sum_{-\infty}^{+\infty}f(q^n
)q^n
\end{eqnarray}
and from $a$ to $\infty$,
\begin{equation}
\int_{a}^{\infty}f(x)d_q x= (1-q)a\sum_{n=1}^{+\infty}f(aq^{-n}
)q^{-n}.
\end{equation}
 Some $q$-functional spaces will be used to establish our result. We begin by putting
\begin{eqnarray}
\R_q &=&\{\pm q^k , k \in \mathbb{Z} \}\cup \{0\},\\
\R_{q,+}&=&\{+ q^k , k \in \mathbb{Z} \}.
\end{eqnarray}
Let $L_{q}^{p}(\R_{q,+})$,  $p \in [1,+\infty[ $ be the space of
functions $f$ such that
\begin{equation}
\parallel f \parallel_{q,p}=\ds(\int_{0}^{\infty}\mid f(x)\mid^{p}d_q x )^{\frac{1}{p}} <
+\infty,
\end{equation}
and for $p = \infty$
\begin{equation}
\parallel f \parallel_{q,\infty} = ess\sup_{x \in \R_{q,+}} \mid f(x)\mid
\end{equation}
Note that for $n \in \mathbb{Z}$ and $a \in \R_q$, we have
\begin{equation}
\int_{0}^{\infty}f(q^n x)d_q x =\frac{1}{q^n
}\int_{0}^{\infty}f(x)d_q x.
\end{equation}
\begin{equation}\label{as}
 \int_{0}^{a}f(q^n x)d_q x =
\frac{1}{q^n }\int_{0}^{aq^n }f(x)d_q x.
\end{equation}

 The $q$-integration by parts is given for suitable function $f$
and $g$ by:
\begin{equation}
\int_{0}^{\infty}f(x)D_{q,x} g(x)d_q x =
\Big[f(x)g(x)\Big]_{0}^{\infty} - \int_{0}^{\infty}D_{q,x}
(f(q^{-1}x))g(x)d_q x.
\end{equation}
Jackson in \cite{jacks} defined the $q$-analogue of the Gamma
function as
\begin{equation}
\Gamma_q (x)=\frac{(q;q)_{\infty}}{(q^{x};q)_{\infty}}(1-q)^{1-x}
\end{equation}
We take the definition of $q$-trigonometric given by T.H.Koornwinder
and R.F.Swarttouw (see \cite{tr}) with simple changes and we write
$q$-cosine and $q$-sinus  as a series of functions
\begin{eqnarray}\label{ba}
\cos(x;q^2
)&=&{}_1\varphi_1(0,q,q^2;(1-q)^2x^2)=\sum_{n=0}^{\infty}(-1)^nb_n(x;q^2)
\\
\sin(x;q^2)&=&(1-q)x{}_1\varphi_1(0,q^3,q^2;(1-q)^2x^2)=\sum_{n=0}^{\infty}(-1)^nc_n(x;q^2)
\end{eqnarray}
where we have put
\begin{eqnarray}
b_n(x;q^2)&=& b_n(1;q^2)x^{2n}=q^{n(n-1)}\frac{(1-q)^{2n}}{(q;q)_{2n}}x^{2n}\\
c_n(x;q^2)&=&c_n(1;q^2)x^{2n+1}=q^{n(n-1)}\frac{(1-q)^{2n+1}}{(q;q)_{2n+1}}x^{2n+1}.
\end{eqnarray}
 The reader will notice that the previous
definition (\ref{ba}) derived from those given in \cite{tr} with
minor change. These functions satisfy \bean
\label{dcos} D_{q,x}\cos(x;q^2)&=& -q^{-1} \sin(qx;q^2),\\
\label{dsin} D_{q,x}\sin(x;q^2)&=& \cos(x;q^2).
 \eean
 and we have the following estimations:
 \bean \label{cos} |\cos(x;q^2)|\leq\frac{1}{(q;q^2)_\infty^2},\\
 \label{sin} | \sin(x;q^2)|\leq\frac{1}{(q;q^2)_\infty^2}.\eean
 We recall the tow
$q$-analogue of the exponential functions \cite{koow}, defined by:
\bean \label{e1} E(x;q)=(-(1-q)x;q)_\infty = \sum_{n=0}^\infty
\frac{(1-q)^n}{(q;q)_n}q^{n(n-1)/2}x^n,\hspace{0.5cm} x\in
\mathbb{R}
\\ \label{e2}
 e(x;q)=\frac{1}{((1-q)x;q)_\infty} = \sum_{n=0}^\infty
\frac{(1-q)^n}{(q;q)_n}x^n, \hspace{0.5cm}|x|<\frac{1}{1-q} . \eean
The function $E(x;q)$ is analytic and $e(z;q)$ is a meromorphic
function on $\mathbb{C}$ having simple poles at $
z=\ds\frac{q^{-m}}{1-q}, m \in \mathbb{N}.$ They satisfy \bean
e(x;q) E(-x;q)=1.\eean

\begin{prop}\label{pri}

1- If $F$ is any $q$-derivative of the function $f$, namely $D_{q,x}
F(x)=f(x)$, continuous at $x=0$, then \bean\label{jx} \int_0^{x}
f(t) d_q t=
  ( F(x) - F(0) ) .\eean
2- For any function $f$ we have \bean D_{q,x} \left[\int_0^{x} f(t)
d_q t \right] =f(x)\eean 3-
  If $G$ is any $q$-derivative of the function $g$, integrable over $(x,\infty)$; $x\geq 0$\\ then:
  \bean
  \int_x^{+ \infty} g(t) d_q t=  -\lim_{b\longrightarrow + \infty}
  ( G(x) - G(b) )
   &=&G(\infty) - G(x)  \eean
 4- For any function $f$ integrable over $(x,+\infty)$, we have
\bean D_{q,x}\left[\int_x^{+ \infty} g(t) d_q t \right]= - g(x)\eean

\end{prop}
\subsection{The $q$-Wronskian:} Let the following $q$-difference
equation: \bean D_{q,x}^2u(x)+a(x)D_{q,x}u(x)+b(x) u(q x)=0
\label{dif1}\eean
\begin{prop}\label{wro} We define the $q$-Wronskian $W(x ; q)$
For the  two solutions $u_1$ and $u_2$  of the $q$-difference
equation (\ref{dif1}) by: \bean \label{www} W(x; q)
(x)&=&u_1(q x)D_{q,x} u_2(x)- u_2(q x) D_{q,x} u_1(x)\label{wr1}\\
&=&\frac{u_1(q x) u_2(x)-u_1(x) u_2(q x)  }{(1-q)x}.
\label{wr2}\eean It satisfies the following $q$-difference equation:
\bean D_{q,x} W(x ;q) +a(x) W(x ;q) =0\label{wr3}.\eean Furthermore:
\bean W(x ;q)=\frac{W(0 ;q)}{\ds\prod_{k=0}^\infty[1+(1-q) q^k x
a(q^k x)]}\label{wr4}\eean
\end{prop}
\begin{proof}
 It easy to see that, if $u_1$ and $u_2$ are tow solutions
of (\ref{dif1}), we have the relation (\ref{wr3}).
\\Using the $q$-derivative definition (\ref{der}), the equation (\ref{wr3}) can be rewritten as:
\bean  W(x ;q) -W(q x ;q) = -(1-q) x a(x) W(x ;q)\eean then \bean
W(q x,q)= \frac{1}{1+(1-q) q x a(q x)} W(q^2 x ;q). \eean So, by
induction we have for $n \in \mathbb{N}$:  \bean W(q^n x ;q)=
\frac{1}{1+(1-q) q^n x a(q^n x) } W(q^{n+1} x ;q).\eean We deduce
that \bean W(x ;q)=\frac{W(0 ;q)}{\ds\prod_{k=0}^\infty[1+(1-q) q^k
x a(q^k x)]}.\eean \end{proof}
 \begin{prop}\label{exx}

  The solution of the $q$-difference equation
\begin{eqnarray}
(E) = \left\{
\begin{array}{lcl}
D_{q,x}^2u(x)+u(q x)= 0    \label{s1}\\\\
  u(0;q^2)=a \\\\
D_qu(0;q^2)=b
\end{array}
\right.
\end{eqnarray}
is given by:
 \bean \label{sol} u(x;q^2) = a \cos(x;q^2) + b
q^{-\frac{1}{2}}sin(q^{\frac{1}{2}}x;q^2)\eean
\end{prop}
\begin{proof}
 Let $u(x) = \ds\sum_{n\geq0}
 a_n x^n$. Then \bean \label{s3}D_{q,x}^2 u(x) =
\sum_{n\geq0} a_{n+2}\frac{1-q^{n+2}}{1-q}\frac{1-q^{n+1}}{1-q} x^n.
\eean If we replace  in (\ref{s1}) we have the following recurrence
relation: \bean \label{rec1}
\frac{1-q^{n+2}}{1-q}\frac{1-q^{n+1}}{1-q}a_{n+2}=q^n a_n .\eean
then\\

  If $n=2p$; $p \in
\mathbb{N}$,  we have \bean q^{2p}
a_{2p}=\frac{1-q^{2p+2}}{1-q}\frac{1-q^{2p+1}}{1-q}a_{2p+2} \eean
so, by induction on $p$, we obtain:
\bean\label{rec2}a_{2p}=(-1)^p\frac{(1-q)^{2p}}{(q;q)_{2p}}
q^{p(p-1)} a_0.\eean
 similarly\\

 if $n=2p+1$, we obtain \bean
\label{rec2}a_{2p+1}=(-1)^p\frac{(1-q)^{2p+1}}{(q;q)_{2p+1}}
q^{p(p-1)} a_1.\eean Using the definition (\ref{ba}), the solution
of (\ref{s1}) is given.\end{proof}
\begin{cor}\label{rr}
For $ x \in \R_q $ we have  \bean \label{un}\cos(q x;q^2)
\cos(q^{\frac{1}{2}}x;q^2)+q^{-\frac{3}{2}}sin(q^{\frac{3}{2}}x;q^2)sin(q
x;q^2)=1\eean which tends to the classical trigonometric relation
\bean cos^2(x)+sin^2(x) =1\eean when $q\rightarrow1^-$
\end{cor}
\begin{proof} Using the relation (\ref{wr3}) and the condition
$a(x)=0$, we have
\begin{equation}
D_{q,x} W(x ;q) =0.
\end{equation}
So,
$$W(x ;q)=\lim_{n\rightarrow+\infty}W(q^n x ;q)=W(0 ;q).$$
\begin{eqnarray*}
 W(0 ;q) &=& \lim_{x\rightarrow 0}W(x ;q)= \lim_{x \rightarrow 0}\left[ \cos(q x;q^2)
\cos(q^{\frac{1}{2}}x;q^2)+ q^{-\frac{3}{2}}\sin(q^{\frac{3}{2}}x
;q^2)\sin(q
x ;q^2) \right ]\\
&=& \lim_{x \rightarrow 0}\frac{1}{(1-q)x}\left[  q^{-\frac{1}{2}}
\cos(x;q^2)\sin(q^{\frac{3}{2}}x;q^2)-
q^{-\frac{1}{2}}  \cos(q x;q^2)\sin(q^{\frac{1}{2}}x;q^2)\right] \\
&=& 1
\end{eqnarray*}

The result follows.
\end{proof}

\section{Asymptotic expansion of solutions at infinity }
In this section, we try to study for $\lambda\longrightarrow\infty $
the asymptotic  expansion of solution $u(x,\lambda^2;q^2)$ of $L_q$
the $q$-difference operator defined by:\bean L_q u(x) =D_{q,x}^2
u(x) -p(x) u(x);\hspace{0.5cm} p(x) \in L^\infty(
\mathbb{R}_{q,+})\cap L^1(\mathbb{R}_{q,+}).\eean In the next we try
to resolve the following $q$-difference problem \bean \label{E} L_q
u(x)=-\lambda^2 u(q x),\hspace{0.5cm} x\in \mathbb{R}_{q,+} ,\lambda
\in \mathbb{R}_{q,+}.\eean
\begin{prop}{\bf (The $q$-Gronwall lemma:)}\\
Let $f$ and $g$ be two positive functions, continuous at 0 and
$q$-integrable over all finite interval of $[0,+\infty[$.\\ We
suppose that \bean f(x)\leq C_q+\int_0^x f(t) g((t) d_qt \eean
where $C_q \in \mathbb{R}_{q,+}$.\\
Then \bean f(x)\leq \frac{C_q}{\ds\prod_{k=0}^\infty[1-(1-q) q^k x
g(q^k x)]}\eean
\end{prop}
\begin{proof}
 Let the following $q$-Jackson integral $$y(x) =\int_0^x
f(t) g(t) d_qt,$$ we have, $$D_{q,x} y(x) =f(x) g(x) \leq[C_q+y(x)]
g(x)$$ then, $$ C_q+y(x) \leq \frac{1}{1-(1-q) x g( x)}(C_q+y(q x)$$
and by induction on $n$, we deduce that $$ C_q+y(q^n x) \leq
\frac{1}{1-(1-q)q^n x g(q^n x)}(C_q+y(q^{n+1} x))$$ then
$$C_q+y(x) \leq \frac{1}{\ds\prod_{k=0}^\infty[1-(1-q) q^k x g(q^k
x)]}(C_q+y(0)).$$
 The fact that $y(0)=0$ leads to the result.\end{proof}
\begin{cor}\label{gro}
Let $f$  be a positive function, continuous at 0 and $q$-integrable
over all finite interval of $[0,+\infty[$. We suppose that there
exist two constants $ C_q $ and $M_q$ in $ \mathbb{R}_{q,+}$, such
that
\begin{equation*}
f(x)\leq C_q + M_q \int_0^x f(t) d_q t.
\end{equation*}
Then we have,
 \bean f(x)\leq C_q e(M_q(1-q)x ;q^2)\eean
 where $e(x;q^2)$ is given by (\ref{e2}).
\end{cor}
\begin{definition}\label{def}
Let $U(x ;q^2)$ and $V(x ; q^2)$ be twice $q$-differentiable
functions. We define $\left[ U , V \right]_{q}$ by:
\begin{equation}
\left[ U , V \right]_{q} = U(q x ;q^2)D_{q,x}V(x ;q^2) - V(q x
;q^2)D_{q,x}U(x ;q^2)
\end{equation}
\end{definition}
\begin{prop}($q$-Green formula)\label{green}\\
For $U(x ;q^2)$ and $V(x ; q^2)$  twice $q$-differentiable
functions, we have
\begin{equation}
D_{q,x}\left[ U , V \right]_{q} = V(q x ;q^2 )L_q U(x;q^2) - U(q x ;
q^2)L_q V(x;q^2)
\end{equation}
\end{prop}

\subsection{$q$-Asymptotic behaviour of $\varphi(x, \lambda^2 ; q^2)$ when $\lambda
\longrightarrow \infty$}For $\lambda \in \R_{q,+}$ let $\varphi(x
,\lambda^2 ; q^2)$ the solution of the following $q$-difference
problem $(E_1)$
\begin{eqnarray} (E_1) = \left\{
\begin{array}{lcl}
L_q U(x,\lambda^2;q^2) = -\lambda^2 U(q x,\lambda^2;q^2) ,    \\\\
 U(0 , \lambda^2 ;q^2) = q^{-1}\sin(q\alpha ;q^2),\\\\
D_q U(0 , \lambda^2 ; q^2) = \cos(q\alpha ; q^2) \quad ,\alpha \in
\R.
\end{array}
\right.
\end{eqnarray}
where $L_q$ is given by (\ref{E}).
\begin{thm}\label{kk}
 Let $p(x)$ in $L_{q}^{\infty}(\R_{q,+})$, then the solution $\varphi(x ,\lambda^2 ; q^2)$ of
$(E_1)$ verifies the following $q$-integral equation:
\begin{eqnarray}\label{pp}
\varphi(x ,\lambda^2; q^2) &=& q^{-1}\sin(q\alpha ; q^2)\cos(\lambda
x ;q^2) + q^{-\frac{1}{2}}\frac{\cos(q \alpha ;
q^2)}{\lambda}\sin(q^{\frac{1}{2}}\lambda x ; q^2) \\ &+&
\frac{1}{\lambda}\int_0^x G(x,y,\lambda^2 ;q^2 )p(y)\varphi(y
,\lambda^2; q^2) d_q y,\nonumber
\end{eqnarray}
where $G(x,y,\lambda^2 ;q^2 )$ is the Green kernel defined by
\begin{equation}
G(x,y,\lambda^2 ;q^2 ) = \cos(q\lambda y;q^2) sin( q^{\frac{1}{2}}
\lambda x;q^2)-sin( q^{\frac{3}{2}}\lambda y;q^2)\cos(\lambda x;q^2)
\end{equation}
\end{thm}
\begin{proof} We begin by resolving the following $q$-homogenous
equation
($E_{1,h}$) \\\\
$(E_{1,h})$: \quad\quad\quad $D_{q,x}^2 U( x ,\lambda^2 ;q^2 ) +
\lambda^2 U(q x ,\lambda^2 ;q^2 ) = 0$.\\\\
For this way, if we use the same steps,  given in proposition
\ref{exx}, we obtain
\begin{equation*}
\varphi_{1,h}(x,\lambda^2 ;q^2)=a \cos(\lambda x;q^2)+ b \frac{
q^{-\frac{1}{2}}}{\lambda} \sin( q^{\frac{1}{2}}\lambda x;q^2) \quad
; a, b \in \R.
\end{equation*}
 Now we are able to give a particular solution $\varphi_p(x ,\lambda^2 ;q^2)$ of $(E_{1})$ .
For deeps, we use the $q$-Method of variation of constant. Hence
 we write $\varphi_p(x ,\lambda^2 ;q^2)$
in the following form
\begin{equation*}
\varphi_p(x ,\lambda^2 ;q^2)=a(x ,\lambda^2 ;q^2) \cos(\lambda
x;q^2)+ b(x ,\lambda^2 ;q^2) \frac{ q^{-\frac{1}{2}}}{\lambda} \sin(
q^{\frac{1}{2}}\lambda x;q^2)
\end{equation*}
  Therefore, if we replace $D_{q,x} \varphi(x ,\lambda^2 ;q^2)$ and
$D_{q,x}^2\varphi(x ,\lambda^2 ;q^2)$ in $(E_1)$, it can be
rewritten in the form
\begin{eqnarray}
D_{q,x}\left[ I_1 \right] + I_2 = p(x) \varphi( x ,\lambda^2 ;q^2),
\end{eqnarray}
where
\begin{equation}
I_1= D_{q,x} a(x ,\lambda^2 ;q^2) \cos( q\lambda x;q^2)+ D_{q,x} b(x
,\lambda^2 ;q^2) \sin(q^\frac{3}{2}\lambda x;q^2)
\end{equation}
and
\begin{equation}
 I_2 = D_{q,x} a(x ,\lambda^2 ;q^2) D_{q,x}[\cos(\lambda x;q^2)] +D_{q,x} b(x
,\lambda^2 ;q^2) D_{q,x}[\sin(q^\frac{1}{2}\lambda x;q^2)].
\end{equation}
 On the other hand, if we use (\ref{dcos}) and (\ref{dsin}), we
 obtain the following system
 \begin{eqnarray}
 \left\{
\begin{array}{lcl}
D_{q,x} a(x ,\lambda^2 ;q^2) \cos( q\lambda x;q^2)+D_{q,x} b(x
,\lambda^2 ;q^2)
sin(q^\frac{3}{2}\lambda x;q^2)&=&0    \\\\
  D_{q,x} a(x ,\lambda^2 ;q^2) D_{q,x}[\cos(\lambda x;q^2)]+D_{q,x} b(x ,\lambda^2 ;q^2)
D_{q,x}[\sin(q^\frac{1}{2}\lambda x;q^2)] &=& p(x) \varphi( x
,\lambda^2 ;q^2)\\\\
\end{array}
\right.
\end{eqnarray}
Hence , \bea D_{q,x} a(x ,\lambda^2 ;q^2)&=&\frac{1}{W(x,\lambda^2
;q^2)}\left|\begin{array}{cc}
  0 & sin(q^\frac{3}{2}\lambda x;q^2) \\
  p(x) \varphi( x ,\lambda^2 ;q^2) & q^\frac{1}{2} \lambda \cos(  q^{\frac{1}{2}}\lambda x;q^2)
\end{array} \right|\\\\
&=& -\frac{p(x) \varphi( x ,\lambda^2 ;q^2)
\sin(q^\frac{3}{2}\lambda x;q^2)} {W(x,\lambda^2 ;q^2 )}, \eea where
$W(x,\lambda^2 ;q^2)$ is given by (\ref{www}). So by
 proposition \ref{pri} and proposition \ref{exx} we obtain
 respectively that
\begin{equation}
W(x,\lambda^2 ;q^2)= \lambda
\end{equation}
and  \bean a(x ,\lambda^2 ;q^2) =-\frac{1}{\lambda}\int_0^x p(y)
\varphi( y ,\lambda^2 ;q^2) \sin(q^\frac{3}{2}\lambda y;q^2)d_q
y.\eean In a similar way, we can show that \bean b(x ,\lambda^2
;q^2)=\frac{1}{\lambda}\int_0^x p(y)\varphi( y ,\lambda^2 ;q^2)
\cos(q\lambda y;q^2)d_q y\eean then the particular solution
$\varphi_p( x ,\lambda^2 ;q^2)$ of $(E_1)$ is given by \bean
\varphi_p( x ,\lambda^2 ;q^2)=\frac{1}{\lambda}\int_0^x p(y)
\varphi( y ,\lambda^2 ;q^2) G(x,y,\lambda^2;q^2) d_q y.\eean where
\bean G(x,y,\lambda^2;q^2)=cos( q \lambda y;q^2) sin(
q^{\frac{1}{2}}\lambda x;q^2)-sin(  q^{\frac{3}{2}}\lambda
y;q^2)\cos( q\lambda x;q^2)\eean
 furthermore, by the fact that $$ \varphi(0,\lambda^2)= a =q^{-1}sin(q\alpha
 ;q^2) $$ and $$ D_q \varphi(0;q^2)=b=cos(q\alpha ;q^2)$$ we can
deduce the result. \end{proof}
\begin{prop}\label{oo}
Let $p(x)$  a boundary function on $\R_{q,+}$. Then $\varphi(x
,\lambda^2 ;q^2)$  verifies:
\begin{enumerate}
\item For $\lambda \in \R_{q,+}$,
\begin{equation}\label{e4}
\varphi(x ,\lambda^2 ;q^2) =\mathcal{O}\left(e(C_\lambda (1-q) x
;q^2)\right)
\end{equation}
where
\begin{equation}
 C_\lambda =\frac { 2\parallel p\parallel_{q,\infty} }{  \mid \lambda \mid
(q;q^2)_\infty^2 }
\end{equation}
and $e( x ; q^2)$ is given by (\ref{e2}).
 \item Additionally, if $\lambda\longrightarrow \infty$, we have
\begin{equation}
\varphi(x ,\lambda^2 ;q^2) =\mathcal{O}(1 ;q^2)
\end{equation}
\end{enumerate}
\end{prop}
\begin{proof}To prove the first result, it suffices to use
 theorem \ref{kk}, (\ref{cos}) and (\ref{sin}). Therefore,
\begin{eqnarray*}\mid \varphi(x ,\lambda^2 ;q^2)\mid & \leq & \frac {
\mid a \mid }{ (q;q^2)_\infty^2 } + \frac {\mid b \mid }{ q \mid
\lambda \mid (q;q^2)_\infty^2 }
+ \frac {2} { \mid \lambda \mid (q;q^2)_\infty^2 } \int_0^x \mid \varphi(y ,\lambda^2 ;q^2)\mid \mid p(y)\mid d_{q}y ,\\
& \leq &\frac { \mid a\mid }{ (q;q^2)_\infty^2 } + \frac { \mid
b\mid }{ q | \lambda | (q;q^2)_\infty^2 }
+ \frac {2} { \mid \lambda \mid (q;q^2)_\infty^2 }\parallel p\parallel_{q,\infty} \int_0^x  \mid \varphi(y ,\lambda^2 ;q^2)\mid d_{q}y,\\
\mid \varphi(x ,\lambda^2 ;q^2)\mid & \leq& A_\lambda +C_\lambda
\int_0^x \mid \varphi(y ,\lambda^2 ;q^2)\mid d_{q}y,
\end{eqnarray*}
where
$$A_\lambda =\frac { \mid a \mid }{ (q;q^2)_\infty^2 } + \frac {\mid b \mid }{ q \mid \lambda \mid (q;q^2)_\infty^2 }$$
and
$$C_\lambda=\frac { \parallel p\parallel_{q,\infty} }{  \mid \lambda \mid (q;q^2)_\infty^2 }$$
 By proposition \ref{gro}, we obtain that
\begin{equation}\label{dd}
\mid \varphi(x ,\lambda^2 ;q^2)\mid \leq A_\lambda e(C_\lambda (1-q)
x ;q^2).
\end{equation}
and the result follows immediately.
\end{proof}
\begin{thm}\label{soll}
 For $p(x)$ in $L_{q}^{\infty} (\R_{q,+})
\cap L^1_{q}(\R_{q,+})$, we have
\begin{enumerate}
\item For $\lambda \geq
\xi > 0$, $\varphi(x ,\lambda^2 ; q^2)$ is an uniformly bounded function.  \\
\item For a large $\lambda$ , we have the following
estimation
\begin{equation}\label{sol}
\varphi(x ,\lambda^2 ; q^2) = \mu(\lambda^2 ; q^2)\cos(\lambda x
;q^2)+ \nu(\lambda^2 ;
q^2)q^{-\frac{1}{2}}\sin(q^{\frac{1}{2}}\lambda x ; q^2) +
\mathcal{O}(1 ; q^2)
\end{equation}
where
\begin{equation}
\mu(\lambda^2 ; q^2)= q^{-1}\sin(q\alpha ;q^2) -
\frac{1}{\lambda}\int_{0}^{\infty}sin( q^{\frac{3}{2}}\lambda
y;q^2)p(y)\varphi(y ,\lambda^2 ; q^2)d_q y
\end{equation}
and
\begin{equation}
\nu(\lambda^2;q^2)=  \frac{\cos(q \alpha ; q^2)}{\lambda} +
\frac{q^{\frac{1}{2}}}{\lambda} \int_{0}^{\infty}\cos(q \lambda y ;
q^2 )p(y)\varphi(y ,\lambda^2; q^2)d_q y.
\end{equation}
\end{enumerate}
\end{thm}
\begin{proof}
 The first result follows immediately by proposition \ref{oo}.\\
proving the second relation, by theorem \ref{kk} we have
\begin{eqnarray*}\label{pp}
\varphi(x ,\lambda^2; q^2) &=& q^{-1}\sin(q\alpha ; q^2)\cos(\lambda
x ;q^2) - q^{-\frac{1}{2}}\frac{\cos(q \alpha ;
q^2)}{\lambda}\sin(q^{\frac{1}{2}}\lambda x ; q^2) \\ &+&
\frac{1}{\lambda}\int_0^{\infty} G(x,y,\lambda^2 ;q^2 )p(y)\varphi(y
,\lambda^2; q^2) d_q y - \frac{1}{\lambda}
\int_{x}^{\infty}G(x,y,\lambda^2 ;q^2 )p(y)\varphi(y ,\lambda^2;
q^2) d_q y
\end{eqnarray*}
taking account of the fact that $p$ in $L_q^1(\R_{q,+})$ and by
(\ref{cos}) , (\ref{sin})
\begin{eqnarray*}
\mid \int_{x}^{\infty}G(x,y,\lambda^2 ;q^2 )p(y)\varphi(y
,\lambda^2; q^2) d_q y\mid &\leq& C_q \int_{x}^{\infty}\mid
p(y)\varphi(y
,\lambda^2; q^2)\mid d_q y \\
&\leq& C_q \parallel \varphi
\parallel_{q,\infty}\int_{0}^{\infty}\mid p(y)\mid d_q y\\ &<& +\infty.
\end{eqnarray*}
leads to the result.
\end{proof}

On the same way, if we note by $\theta(x,\lambda^2 ; q^2)$ the
solution of the  $q$-difference problem
\begin{eqnarray}
(E_2) = \left\{
\begin{array}{lcl}
L_q U = -\lambda^2 U     \\\\
 U(0 , \lambda^2 ;q^2) = q^{\frac{1}{2}}\cos(q^{\frac{1}{2}}\alpha ;q^2)\\\\
D_q U(0 , \lambda^2 ; q^2) = -\sin(q^{\frac{3}{2}}\alpha ; q^2)
\quad ,\alpha \in \C
\end{array}
\right.
\end{eqnarray}
We show that
\begin{equation}
\theta(x,\lambda^2 ;q^2)= \mu_1 (\lambda^2 ;q^2)\cos(\lambda x ;q^2)
+ \nu_1 (\lambda^2 ;q^2) q^{-\frac{1}{2}}\sin(q^{\frac{1}{2}}\lambda
x ; q^2) + \mathcal{O}(1 ; q^2)
\end{equation}
where
\begin{equation}
\mu_1(\lambda^2 ; q^2)= q^{\frac{1}{2}}\cos(q^{\frac{1}{2}}\alpha
;q^2) - \frac{1}{\lambda}\int_{0}^{\infty}sin(
q^{\frac{3}{2}}\lambda y;q^2)p(y)\varphi(y ,\lambda^2 ; q^2)d_q y
\end{equation}
and
\begin{equation}
\nu_1(\lambda^2;q^2)= - q^{\frac{1}{2}}\frac{\cos(q \alpha ;
q^2)}{\lambda} - q^{\frac{1}{2}}\frac{\sin(q^{\frac{3}{2}}\alpha ;
q^2)}{\lambda} \int_{0}^{\infty}\cos(q \lambda y ; q^2
)p(y)\varphi(y ,\lambda^2; q^2)d_q y.
\end{equation}
\begin{thm}
For $\lambda$ in $\R_{q,+}$, we have:
\begin{equation}
\mu(\lambda^2 ; q^2)\nu_1(\lambda^2 ; q^2)- \nu(\lambda^2 ;
q^2)\mu_1(\lambda^2 ; q^2) = \frac{1}{q^{\frac{1}{2}} \lambda}.
\end{equation}
\end{thm}
\begin{proof} Using  theorem \ref{soll}, we obtain that
\begin{equation*}
\varphi(x ,\lambda^2 ; q^2) = \mu(\lambda^2 ; q^2)\cos(\lambda x
;q^2)+ \nu(\lambda^2 ;
q^2)q^{-\frac{1}{2}}\sin(q^{\frac{1}{2}}\lambda x ; q^2) +
\mathcal{O}(1 ; q^2)
\end{equation*}
and
\begin{equation*}
\theta(x,\lambda^2 ;q^2)= \mu_1 (\lambda^2 ;q^2)\cos(\lambda x ;q^2)
+ \nu_1 (\lambda^2 ;q^2) q^{-\frac{1}{2}}\sin(q^{\frac{1}{2}}\lambda
x ; q^2) + \mathcal{O}(1 ; q^2).
\end{equation*}
We can deduce that
\begin{equation*}
D_{q,x}\varphi(x ,\lambda^2 ; q^2) = - q^{-1}\lambda\mu(\lambda^2 ;
q^2)\sin(q\lambda x ;q^2)+ \lambda \nu(\lambda^2 ;
q^2)\cos(q^{\frac{1}{2}}\lambda x ; q^2) + \mathcal{O}(1 ; q^2)
\end{equation*}
and
\begin{equation*}
D_{q,x}\theta(x,\lambda^2 ;q^2)= - \lambda q^{-1}\mu_1 (\lambda^2
;q^2)\sin(q\lambda x ;q^2) +  \lambda \nu_1 (\lambda^2 ;q^2)
\cos(q^{\frac{1}{2}}\lambda x ; q^2) + \mathcal{O}(1 ; q^2),
\end{equation*}
then, the use of definition \ref{def} leads to
\begin{eqnarray*}
\left[ \varphi , \theta \right]_{q}(x) &=& \varphi(q x
;q^2)D_{q,x}\theta(x ;q^2) - \theta(q x
;q^2)D_{q,x}\varphi(x ;q^2)\\
&=& \lambda \big[\nu(\lambda^2 ; q^2)\mu_1(\lambda^2 ; q^2)-
\mu(\lambda^2 ; q^2)\nu_1(\lambda^2 ; q^2)\big]\big[
\cos(q\lambda x ; q^2)\cos(q^{\frac{1}{2}}\lambda x ; q^2) \\
&+& q^{-\frac{3}{2}}\sin(q^{\frac{3}{2}}\lambda x ; q^2)\sin(q
\lambda x ; q^2)\big] + \mathcal{O}(1 ; q^2).
\end{eqnarray*}
In the other side, the fact that $\varphi$ and $\theta$ are
solutions of (E) and   by proposition \ref{green} we can deduce that
\begin{equation*}
D_{q,x}\left[ \varphi , \theta \right]_{q}(x) = 0.
\end{equation*}
then from proposition \ref{pri}
\begin{eqnarray*}
\left[ \varphi , \theta \right]_{q}(x) &=& \left[ \varphi , \theta
\right]_{q}(0)= q^{\frac{1}{2}}\cos(q^{\frac{1}{2}}\alpha
;q^2)\cos(q\alpha ; q^2)+ q^{-1}\sin(q\alpha
;q^2)\sin(q^{\frac{3}{2}}\alpha ; q^2)\\
&=& q^{\frac{1}{2}}
\end{eqnarray*}
the result follows.
\end{proof}
\section{Asymptotic behaviour of $j_\alpha(\lambda x;q^2)$:} In
this section, our objective is to establish; using the method of
variation of constant given in the last section; the asymptotic
expansion of $j_\alpha(\lambda x;q^2)$ when $\lambda \longrightarrow
+\infty$.\\ we recall some properties given in \cite{fitham}: For
$\alpha>-\frac{1}{2}$, the $q$-Bessel function is defined by:
\begin{equation}\label{Jj}
 j_\alpha(x;q^2)=\Gamma_{q^2}(\alpha+1) \frac{q^\alpha
 (1+q)^\alpha}{x^\alpha}J_\alpha((1-q)x;q^2)
\end{equation}where
$J_\alpha(x;q^2) $ is the $q$-Bessel Han Exton \cite{swart}, defined
by \bean J_\alpha(z;q) =\left(\frac{z}{1-q}\right)^\alpha
\sum_{k=0}^\infty \frac{(-1)^k q^{\frac{k(k-1)}{2}}q^k}
{\Gamma_q(k+1)\Gamma_q(\alpha + k + 1)}\left(\frac
{z}{1-q}\right)^{2k}.\eean
  This function satisfies  the following relations
   \bean j_{-\frac{1}{2}}(x;q^2)&=& \cos(x;q^2)\\
   j_{\frac{1}{2}}(x;q^2)&=&\frac{1}{x}\sin(x;q^2).\eean
\begin{prop}
   The function $j_\alpha(\lambda x;q^2)$; $\lambda$ being complex; is the solution of
   the following   $q$-difference problem:
\begin{eqnarray}
(E) = \left\{
\begin{array}{lcl}
\Delta_{q,\alpha} y(x)+\lambda^2 y(x) = 0     \\\\
   y(0)=1 \\\\
D_q y(0)=0
\end{array}
\right.
\end{eqnarray}
where $\Delta_{q,\alpha}$ is the $q$-Bessel operator, defined by
\bean  \Delta_{q,\alpha}f(x) &=&\frac{1}{x^{2\alpha+1}}D_{q,x}\left[
x^{2\alpha+1} D_{q ,x}f \right](q^{-1}x),\\ & =&
q^{2\alpha+1}\Delta_{q,x} f(x) +\frac{1- q^{2\alpha+1}}{(1-q) q^{-1}
x} D_{q,x} f(q^{-1}x)\eean and \bean \Delta_{q,x} f(x)= (D_{q,x}^2
f)(q^{-1}x)\eean
\end{prop}
\begin{cor}\label{ss} For $x\in \mathbb{R}_q$ and  $\frac{\log(1-q)}{\log q } \in
\mathbb{Z} $, we have the following estimation
 \bean |j_\alpha(x;q^2)|&\leq& \frac{1}{(q;q^2)^2_\infty}
 \eean
  \end{cor}

Now, we consider the following $q$-Bessel equation given by: \\\\
$(E)$ \quad $\ds
D_{q,x}^2y(x,\lambda^2;q^2)+\ds\frac{\lambda^2}{q^{2\alpha+1}}
y(q x,\lambda^2;q^2)= -\ds\frac{1-q^{2\alpha+1}}{q^{2\alpha+1}(1-q)x}D_{q,x} y(x,\lambda^2;q^2)$.\\\\
Let
$y_h(x,\lambda^2;q^2)=\ds\sum_{n \geq 0} a_n(\alpha,\lambda^2;q^2) x^n$ the homogeneous solution of $(E_h)$ given by \\
 $(E_h)$\quad
$\ds D_{q,x}^2y(x,\lambda^2;q^2)+\ds\frac{\lambda^2}{q^{2\alpha+1}}
y(q x,\lambda^2;q^2)= 0$.\\\\ Then, we have \bea  D_{q,x}^2
y(x,\lambda^2;q^2) =\sum_{n \geq 2}\frac{(1-q^n)
(1-q^{n-1})}{(1-q)^2} a_n x^{n-2} =\sum_{n \geq 0}\frac{(1-q^{n+2})
(1-q^{n+1})}{(1-q)^2} a_{n+2} x^n.\eea By identification, we obtain
the following recurrence relation: \bean \frac{(1-q^{n+2})
(1-q^{n+1})}{(1-q)^2} a_{n+2} =- \frac{\lambda^2 q^n
}{q^{2\alpha+1}}a_n.\eean If we proceed in a similar way of
proposition \ref{exx}, we can deduce easily that the homogeneous
solution $y_h (x, \lambda^2 ;q^2)$ is given by \bean y_h (x,
\lambda^2 ;q^2) =a_0 \cos( q^{-\alpha-\frac{1}{2}}\lambda x;q^2) +
a_1\frac{ q^{-\alpha-1}}{\lambda}\sin (q^{-\alpha} \lambda  x ;q^2)
\eean where $a_0 , a_1 $ are constants in
$\R$.\\
Now, we give a particular solution $y_p(x, \lambda^2 ;q^2)$ of (E)
in the form  \bean y_p(x, \lambda^2 ;q^2)= a_\alpha(x, \lambda^2
;q^2) \cos ( q^{-\alpha-\frac{1}{2}}\lambda x;q^2) + b_\alpha(x,
\lambda^2 ;q^2)\frac{ q^{-\alpha-1}}{\lambda} \sin(q^{-\alpha}
\lambda x ;q^2). \eean The $q$-wronskian is given by \bea
W(x,\lambda^2 ;q^2)=D_{q,x}\left[ \cos(
q^{-\alpha-\frac{1}{2}}\lambda x;q^2)\right]\frac{
q^{-\alpha-1}}{\lambda}\sin(q^{-\alpha} \lambda  x ;q^2)\\ - \cos (
q^{-\alpha-\frac{1}{2}}\lambda x;q^2) D_{q,x}\left[\frac{
q^{-\alpha-1}}{\lambda}\sin (q^{-\alpha} \lambda  x ;q^2)\right],
\eea
 thus, using corollary \ref{rr} we can show that
$$D_{q,x} W(x,\lambda^2 ;q^2) =0,$$ therefore
\begin{eqnarray}
W(x , \lambda^2 ;q^2) = - q^{-2\alpha -1}\lambda.
\end{eqnarray}
Using the $q$-method of variation of constant (given in theorem
\ref{kk}) and proposition \ref{pri} we deduce that
\begin{eqnarray}
a_\alpha(x, \lambda^2 ;q^2)&=&
\frac{q^{-\alpha-1}}{\lambda}\int_x^{+\infty} \frac{1-q^{2\alpha
+1}}{1-q}\frac{1}{t}D_{q,t} y(t, \lambda^2 ;q^2) \sin(q^{-\alpha +1}\lambda t ;q^2) d_q t\\
b_\alpha(x, \lambda^2 ;q^2)&=&-
\frac{1}{\lambda}\int_{x}^{+\infty}\frac{1-q^{2\alpha
+1}}{1-q}\frac{1}{t}D_{q,t} y(t, \lambda^2 ;q^2) \cos(q^{-\alpha
+\frac{1}{2}}\lambda t ;q^2) d_q t
\end{eqnarray}
this leads to the following result
\begin{prop}For $x , \lambda \in \R_{q,+}$, the solution $y(x , \lambda^2 ;q^2)$ of (E) is given by:
\begin{eqnarray*}
 y(x ,\lambda^2 ;q^2) &=& a \cos(q^{-\alpha - \frac{1}{2}}\lambda x
;q^2) + \frac{b}{\lambda}q^{-\alpha -1}\sin(q^{- \alpha}\lambda x ;q^2) \\
&+& \frac{q^{-\alpha -1}}{\lambda}\int_{x}^{+\infty}G_\alpha (t , x
,\lambda ;q^2)\frac{1-q^{2\alpha +1}}{1-q}\frac{1}{t}D_{q,t}y(t ,
\lambda^2 ;q^2)d_qt
\end{eqnarray*}
where
\begin{equation}
G_\alpha (t , x ,\lambda ;q^2)=\cos(q^{-\alpha - \frac{1}{2}}\lambda
x ;q^2)\sin(q^{-\alpha + 1}\lambda t)- \cos(q^{-\alpha +
\frac{1}{2}}\lambda t ;q^2)\sin(q^{-\alpha} \lambda x ;q^2)
\end{equation}
\end{prop}
On the other hand   $j_{\alpha}(\lambda x;q^2)$ is the unique
solution of (E) with  initial conditions
\begin{eqnarray*}
j_{\alpha}(0;q^2)&=&1 \\
D_q j_{\alpha}(0;q^2)&=&0.
\end{eqnarray*}
Therefore,  we can write $j_{\alpha}(\lambda x;q^2)$ as the
following form
\begin{thm}\label{res} For $\lambda \in \R_{q,+}$ and $\alpha > -\frac{1}{2}$, we have
\begin{enumerate}
\item
\begin{eqnarray}
j_{\alpha}(\lambda x;q^2) = \cos( q^{- \alpha - \frac{1}{2}}\lambda
x;q^2) + R_{q,\alpha} (x,\lambda^2 ),
\end{eqnarray}
where
\begin{eqnarray*}
  R_{q,\alpha} (x,\lambda^2 )&=& -\frac{1}{\lambda}\frac{1-q^{2\alpha +
1}}{1-q}\int^{+\infty}_{x}G_\alpha (t , x ,\lambda ;q^2) \frac{1}{t}
D_{q,t} j_{\alpha}(\lambda t;q^2) d_q t
\end{eqnarray*}
\item Additionally, $R_{q,\alpha} (x,\lambda^2)$ tend to $0$ when
$\lambda$ tend to $+\infty$.
\end{enumerate}
\end{thm}
\begin{proof}To prove 2. it suffices to use (\ref{cos}),
(\ref{sin}) and corollary\ref{ss}, then
\begin{eqnarray*}
| R_{q,\alpha} (x,\lambda^2 )|&\leq&
\frac{1}{\lambda}\frac{1-q^{2\alpha + 1}}{1-q}\frac{2}{(q;q^2
)_{\infty}^2}\int^{+\infty}_{x}\mid\frac{D_{q,t} j_{\alpha}(\lambda
t;q^2)}{t}\mid d_q t \\
&\leq& \frac{1-q^{2\alpha + 1}}{1-q}\frac{2}{(q;q^2
)_{\infty}^2}\int^{+\infty}_{\lambda
x}\mid\frac{D_{q,t} j_{\alpha}(t;q^2)}{t}\mid d_q t \\
&\leq& \frac{1-q^{2\alpha + 1}}{1-q}\frac{2}{(q;q^2
)_{\infty}^2}\int^{+\infty}_{\lambda
x}\frac{|j_\alpha (t;q^2)- j_\alpha (qt;q^2)|}{(1-q)t^2}d_qt\\
&\leq& \frac{1-q^{2\alpha + 1}}{(1-q)^2}\big[\frac{2}{(q;q^2
)_{\infty}^2}\big]^2
\int^{+\infty}_{\lambda x}\frac{d_q t}{t^2} \\
&\leq& \frac{C_q }{\lambda x}\longrightarrow 0 \quad,\lambda
\longrightarrow \infty
\end{eqnarray*}
where
\begin{equation}
 C_q =\frac{1-q^{2\alpha + 1}}{q}\big[ \frac{2}{(1-q)(q;q^2
)_{\infty}^2}\big]
\end{equation}
and the result follows immediately.
\end{proof}
\subsection{Application}
We recall the $q$-equality of Weber integral  study in \cite{fitham}.\\
For $a , \lambda $ in $\R_{q,+}$ and $\alpha >-1 $ we have
\begin{equation}
\frac{1}{A_{\alpha}}\int_{0}^{\infty}e(- a^2 x^2 ;q^2)j_\alpha
(\lambda x ;q^2)x^{2 \alpha +1}d_q x = \frac{1}{a^{2 \alpha
+2}}e(-\frac{\lambda^2}{a^2 (1+q)^2 } ;q^2)
\end{equation}
where
\begin{equation}
A_\alpha = \int_{0}^{\infty}\frac{x^{2 \alpha + 1}}{(-(1-q^2 )x^2
;q^2)}d_q x
\end{equation}
and if we take $a = \sqrt{t}$ we obtain
\begin{equation}
E_\alpha (t , \lambda^2 ;q^2) = \int_{0}^{\infty}e(-q^{-1}t x^2
;q^2)j_\alpha (\lambda x ;q^2)x^{2 \alpha +1}d_q x
\end{equation}
where
\begin{equation}
E_\alpha (t , \lambda^2 ;q^2)= \frac{A_\alpha}{t^{\alpha
+1}}e(-\frac{\lambda^2}{(1+q)^2 t } ;q^2)
\end{equation}
\begin{prop}For $\alpha > -1$ and $\lambda , t \in \R_{q,+}$, the heat kernel $E_\alpha (t , \lambda ;q^2)$ has the
following behaviour:
\begin{equation}
E_\alpha (t , \lambda^2 ;q^2) = \int_{0}^{\infty}e(-q^{-1}t x^2
;q^2)\cos( q^{- \alpha - \frac{1}{2}}\lambda x;q^2) x^{2 \alpha
+1}d_q x + \Theta_\alpha (\lambda , t ;q^2)
\end{equation}
where
\begin{equation}
\Theta_\alpha (\lambda , t ;q^2) \longrightarrow 0 \quad , \lambda
\rightarrow \infty
\end{equation}
\end{prop}
\begin{proof}
the result follows by proposition \ref{res} and the fact that
\begin{eqnarray*}
\mid \int_{0}^{\infty}e(-q^{-1}t x^2 ;q^2)R_q (\lambda^2 ,x) x^{2
\alpha +1}d_q x \mid &\leq&
\frac{C_q}{\lambda^2}\int_{0}^{\infty}e(-q^{-1}t x^2 ;q^2)x^{2
\alpha }d_q x\\
&=&\frac{C_q}{\lambda^2}(1-q)\sum_{-\infty}^{+\infty}\frac{q^{(2\alpha+1)
k}}{(-q^{-1}(1-q^2)tq^{2k} ; q^2)_\infty}\\
&=&\frac{C_q}{\lambda^2}(1-q)\sum_{-\infty}^{+\infty}\frac{q^{2\beta
k}}{(aq^{2k} ; q^2)_\infty}
\end{eqnarray*}
where $\beta = \alpha +\frac{1}{2}$ and $a = -q^{-1}(1-q^2)t$ . The
use of Ramanujan's sum (see \cite{ismail}) leads to
\begin{eqnarray*}
\mid \int_{0}^{\infty}e(-q^{-1}t x^2 ;q^2)R_q (\lambda^2 ,x) x^{2
\alpha +1}d_q x \mid &\leq&
\frac{C_q}{\lambda^2}B_\alpha(t;q^2)\longrightarrow 0 \quad, \lambda
\rightarrow \infty
\end{eqnarray*}
where
\begin{equation}
B_\alpha(t;q^2)= \frac{(-q^{2\alpha}t(1-q^2),
q^{2-\alpha}t^{-1}(1-q^2)^{-1} , q^2 ; q^2)_\infty}{(q^{2\alpha+1} ,
-q^{-1}t(1-q^2) ,-q^3t^{-1}(1-q^2)^{-1} ; q^2)_\infty}.
\end{equation}
\end{proof}

\end{document}